\makeatletter
\def\ps@wanda{%
   \renewcommand{\@oddhead}{\begin{minipage}{13cm}
        \footnotesize\itshape  Appeared in slightly different form in: \\{\rm Journal of Physics A: Mathematical and General {\bf 35} (2002) 5701-5709} 
                            \end{minipage}
                                }%
   \renewcommand{\@evenhead}{}%
   \renewcommand{\@oddfoot}{\hfil\footnotesize\textsf{\bfseries\thepage}\hfil}%
   \renewcommand{\@evenfoot}{}%
}%
\makeatother
\documentclass[10pt]{iopart}
\usepackage{iopams,amsthm}  

\theoremstyle{plain}
\newtheorem{theorem}{Theorem}

\begin{document}

\title[Simple diamagnetic monotonicities]{Simple diamagnetic monotonicities 
for Schr\"odinger operators with  inhomogeneous magnetic fields of constant direction}

\author{Hajo Leschke, Rainer Ruder and Simone Warzel}

\address{Institut f\"ur Theoretische Physik, Universit\"at Erlangen--N\"urnberg, Staudtstr.~7, 91058 Erlangen, Germany}
\ead{hajo.leschke@physik.uni-erlangen.de}

\begin{abstract}
  Under certain simplifying conditions 
  we detect monotonicity properties of the ground-state energy and the 
  canonical-equilibrium density matrix 
  of a spinless charged
  particle in the Euclidean plane subject to a perpendicular, possibly inhomogeneous magnetic field 
  and an additional scalar potential. 
  Firstly, we point out a simple condition   
  warranting that the ground-state energy does not decrease when the magnetic field and/or the potential is 
  increased pointwise.
  Secondly, we consider the case in which both the magnetic field and the potential 
  are constant along one direction in the plane
  and give a genuine path-integral argument for corresponding monotonicities of 
  the density-matrix diagonal and the absolute value of certain off-diagonals. 
  Our results complement to some degree
  results of M. Loss and B. Thaller [{\it Commun.\ Math.\ Phys.\ }{\bf 186} (1997) 95] and L. Erd\H{o}s 
  [{\it J.\ Math.\ Phys.\ }{\bf 38} (1997) 1289].  
\end{abstract}
\vspace*{-0.5cm}
\pacs{75.20.-g, 03.65.Ge
  \\[0.2cm]
  Dedicated to John R. Klauder on the occasion of his $ 70^{\rm th} $ birthday}


\thispagestyle{wanda}
\section{Introduction}
In non-relativistic classical and quantum statistical mechanics of thermal equilibrium 
the ground-state energy and, more generally, 
the free energy of the considered particle system do not decrease when its external potential and/or 
interaction potential is increased. 
This monotonicity property continues to hold in the presence of a fixed external, 
possibly inhomogeneous magnetic field, a fact which is 
trivial in the classical situation, because here
the magnetic field can be eliminated in accordance with  
the Bohr/van Leeuwen theorem \cite{M} on the non-existence of diamagnetism in classical physics.
In the non-trivial quantum case a monotonicity with respect to the magnetic field analogous to that in the potentials 
does not hold in such generality.
Only few results are available. The most general one is due to the so-called diamagnetic inequality 
which implies that the free energy of a system of  spinless distinguishable particles or of spinless bosons 
within an arbitrary magnetic field is not smaller than without \cite{S76,SFI,GP2,R}. 
More generally, for such a system the absolute value of its canonical-equilibrium density matrix, that is,
of the position representation of the underlying ``Boltzmann operator'' (in other words: heat kernel) 
does not decrease when the magnetic field is
switched off. 
As observed by E.~Nelson, this is most easily obtained 
by applying the triangle inequality to the Feynman-Kac(-It\^{o}) path-integral representation of the density matrix \cite{S77,SFI,R}.  
A drawback of the diamagnetic inequality is that it only enables one to compare the two extreme situations 
with or without magnetic field.  
It does not supply control when a magnetic field is changed (pointwise) from a given non-zero value to another non-zero one. 
\emph{Diamagnetic monotonicities} in this more general sense are in fact much harder to obtain and even wrong in general, already 
for one-particle systems. 
This has been nicely discussed by L.~Erd\H{o}s \cite{E97} in terms of illuminating examples and counterexamples 
for one-particle Schr\"odinger and Pauli operators. 
In deriving some of his results, Erd\H{o}s was inspired by an interesting diamagnetic-monotonicity result   
of Loss and Thaller \cite{LT97} for the density matrix of a spinless electrically 
charged particle subject exclusively to an inhomogeneous magnetic
field of constant direction or, what amounts to the same, 
of a particle confined to a plane perpendicular to that field. 
(For a more precise statement, see (\ref{eq:LT}) below.)
 
The present note is inspired by both works \cite{LT97,E97}. 
Given the Schr\"odinger operator for a 
spinless charged particle in the Euclidean plane 
subject to a perpendicular and continuously differentiable but otherwise arbitrary  
magnetic field and to an additional (scalar) potential, we are going to present two results 
which to a certain extent complement
results in \cite{LT97,E97} and provide partial positive answers to the open problem \cite[II.12]{E97}. 
We first point out a simple sufficient but, in general, somewhat implicit condition 
for the ground-state energy not to decrease when the magnetic field and/or the potential is increased pointwise.  
Secondly, a rather direct genuine path-integral argument yields a diamagnetic monotonicity for density matrices 
under the simplifying condition that both the magnetic field and the potential are 
constant along (at least) one direction in the plane. 
As a by-product, the estimate
of Loss and Thaller \cite{LT97} is sharpened for these special fields.
The validity of both results and the simplicity of their proofs heavily rely 
on the fact that under the mentioned conditions the term 
in the Schr\"odinger operator which is linear in the canonical momentum 
can easily be brought under control. 

In the remainder of the Introduction we set the stage and fix our basic notation. 
We consider a spinless charged particle in the  Euclidean plane $ \mathbb{R}^2 $ and 
choose physical units where the mass and the charge of the particle as well as Planck's constant (divided by $ 2 \pi $) 
are all equal to one. 
The corresponding quantum system is characterized by a \emph{Schr\"odinger operator}, or standard Hamiltonian, 
informally given by the expression
\begin{equation}\label{eq:1}
  H({b},{v}) := \case{1}{2} \big(\bi{P} - \bi{{a}}(\bi{Q}) \big)^2 + {v}(\bi{Q})
\end{equation}
where $ \bi{Q}=(Q_1,Q_2) $  and $ \bi{P}=(P_1,P_2) $ 
denote the usual two-component vector operators of position and canonical momentum, respectively. 
Since we are interested in gauge-independent quantities only, we take the liberty to slightly abuse 
the notation in that the operator (\ref{eq:1}) actually depends on the vector potential $ \bi{{a}} = ({a}_1, {a}_2 ) $ 
(and the scalar potential $ v $)
and not just on the magnetic field $ {b}: \mathbb{R}^2 \to \mathbb{R} $ related to $ \bi{a} $ by 
$ {b}(\bi{x}) = \partial {a}_2(\bi{x})/ \partial x_1 - \partial {a}_1(\bi{x}) / \partial x_2 $, where 
$ (x_1, x_2 ) $ is the pair of Cartesian co-ordinates of the point $ \bi{x} \in \mathbb{R}^2 $. 
We write $ \bi{x} \cdot \bi{y} := x_1 y_1 + x_2 y_2 $ for the Euclidean scalar product of 
$ \bi{x} $, $ \bi{y} \in \mathbb{R}^2 $ and also $ \bi{x}^2 := \bi{x} \cdot \bi{x} $ and $ | \bi{x} | := \sqrt{\bi{x}^2} $.
To avoid technical complications, we will assume throughout that $ \bi{a}: \mathbb{R}^2 \to \mathbb{R}^2 $ 
is continuously differentiable and $ {v}: \mathbb{R}^2 \to \mathbb{R} $ 
is continuous and bounded from below. This in particular implies that $ H({b},{v}) $ 
can be defined straightforwardly and uniquely as a self-adjoint bounded below operator
acting on a dense domain in the separable Hilbert space $ {\rm L}^2(\mathbb{R}^2) $ of 
all (equivalence classes of) Lebesgue square-integrable complex-valued functions on the plane
with the usual scalar product $ \langle \varphi , \, \psi \rangle := \int_{\mathbb{R}^2} d^2 x 
\; \varphi(\bi{x})^* \, \psi(\bi{x}) $, for $ \varphi $, $ \psi \in {\rm L}^2(\mathbb{R}^2) $.

The simplest quantity we are going to look at is the \emph{ground-state energy} of $  H({b},{v}) $, that is, 
the bottom of its spectrum, in symbols
\begin{equation}
  e_0({b},{v}) := {\rm inf \; spec} \, H({b},{v}) 
  = \inf_{\langle \psi , \, \psi \rangle = 1 } \, 
  \langle \psi , \,  H({b},{v}) \, \psi \rangle.
\end{equation}
Here the second infimum is taken over all normalized wave-functions $ \psi $ 
in the domain of $  H({b},{v}) $.
The other quantity we will deal with is the canonical-equilibrium \emph{density matrix} of $  H({b},{v}) $, that is, 
$ \langle \bi{x} | \exp[ - \beta H({b}, {v} ) ] | \bi{y} \rangle $.
It is defined as the in $ \bi{x} $, $ \bi{y} \in \mathbb{R}^2 $ jointly continuous Hermitian integral
kernel of the self-adjoint, non-negative and bounded ``Boltzmann operator'' $  \exp[ - \beta H({b}, {v} ) ] $ 
corresponding to inverse temperature $ \beta > 0 $. 
In particular, the image $ \exp[ - \beta H({b}, {v} ) ] \, \psi \in {\rm L}^2(\mathbb{R}^2) $ of an arbitrary $ \psi \in {\rm L}^2(\mathbb{R}^2) $ has a continuous representative given by the function $ \bi{x} \mapsto \int_{\mathbb{R}^2} d^2 y \,  \langle \bi{x} | \exp[ - \beta H({b}, {v} ) ] | \bi{y} \rangle \, \psi(\bi{y}) $ and 
the density-matrix diagonal $ \langle \bi{x} | \exp[ - \beta H({b}, {v} ) ] | \bi{x} \rangle $ is a well-defined, non-negative and continuous function of $  \bi{x} \in \mathbb{R}^2 $.
For the precise definition of $ H({b}, {v} ) $ and the existence of a 
continuous integral kernel of the corresponding ``Boltzmann operator'' (under much weaker assumptions on $ {b} $ and $ { v} $), 
see \cite{CFKS,SFI,BHL00,BML02}. 

\section{Monotonicity of ground-state energies}
The following two facts, which we recall from \cite{E97}, show 
that diamagnetic monotonicity fails in general already for ground-state energies.
\begin{description}
  \item{{\bf Fact 1} (\cite[II.10]{E97}):}
    There exists a centrally symmetric potential $ \hat {v} $ with the function $ | \bi{x} | \mapsto \hat {v}(|\bi{x}|) $ 
    being decreasing in some neighbourhood of some $ \varrho > 0 $. Moreover, there exist  
    two constant magnetic fields $ {b} $ and $ \hat {b} $ such that $  0< {b} < \hat {b} $ but
    $ e_0({b},\hat {v}) >  e_0(\hat {b},\hat {v}) $. 
   \item{{\bf Fact 2} (\cite[II.11]{E97}):}
     There exist two inhomogeneous centrally symmetric magnetic fields $ {b} $ and $ \hat {b} $ such that 
     $ 0 \leq {b}(|\bi{x}|) \leq \hat {b}(|\bi{x}|)  $ for all $ \bi{x} \in \mathbb{R}^2 $ but 
     $ e_0({b},0) >  e_0(\hat {b}, 0) $. 
\end{description}
In both situations the ground-state wave-function $ \hat\psi_0 $ of $ H(\hat {b},\hat {v}) $,
in the \emph{Poincar\'e gauge}
\begin{equation}\label{eq:Poncaregauge}
  \hat\bi{ {a}}(\bi{x}) := \left( - x_2 , x_1 \right) \; 
    \int_0^1 d \xi \, \xi \; \hat {b}\left(\xi \bi{x}\right),
\end{equation}
is an eigenfunction of the canonical angular-momentum operator $ L_3 := Q_1 P_2 - Q_2 P_1  $ 
with a non-zero eigenvalue. Hence $ \hat\psi_0 $ is not real-valued. This motivates 
the additional assumption in

\begin{theorem} \label{Thm:1}
Let the Schr\"odinger operator $ H(\hat {b},\hat {v}) $ in the Poincar\'e gauge (\ref{eq:Poncaregauge})
possess a real-valued ground-state wave-function, 
that is,
$  e_0(\hat {b}, \hat {v}) =  \langle \hat\psi_0 , \, H(\hat {b},\hat {v}) \, \hat\psi_0 \rangle $ 
for some real-valued $ \hat\psi_0 $ in the domain of $ H(\hat {b},\hat {v})$ 
with $ \langle \hat\psi_0 , \, \hat\psi_0 \rangle = 1 $.
Then the pointwise inequalities $  | {b}(\bi{x}) |\leq \hat {b}(\bi{x})  $ and $ {v}(\bi{x}) \leq \hat {v}(\bi{x}) $ 
for all $ \bi{x} \in \mathbb{R}^2 $
imply the inequality 
\begin{equation}\label{eq:Thm1}
 e_0({b}, {v}) \leq e_0(\hat {b}, \hat {v})
\end{equation}
for the corresponding ground-state energies.
\end{theorem}
\noindent
Before giving a simple proof of the theorem at the end of the section, four remarks are in order:
\begin{itemize}
  \item
    An unpleasant feature of Theorem~1 is that the additional assumption uses the
    special gauge (\ref{eq:Poncaregauge}) and therefore is not gauge independent. However,
    this gauge is often convenient, in particular, if $ \hat {b} $ and $ \hat {v} $ are centrally symmetric, see below.
    For other gauges the existence of a real-valued ground-state wave-function $ \hat\psi_0 $ 
    is in general not sufficient to imply the monotonicity (\ref{eq:Thm1}). 
    Conversely, even in the gauge (\ref{eq:Poncaregauge}) real-valuedness of a $ \hat\psi_0 $ is not 
    necessary for (\ref{eq:Thm1}) to hold
    as the diamagnetic inequality $ e_0(0, \hat {v}) \leq e_0(\hat {b}, \hat {v}) $ illustrates \cite{S76}.
    (See also the Introduction and the first remark following Theorem~\ref{Thm2} below.)
  \item
    If $  e_0(\hat {b}, \hat {v}) $ does not belong to the point spectrum of $ H(\hat {b},\hat {v}) $, equivalently,
    if there is no square-integrable ground-state wave-function $ \hat \psi_0 $, assertion (\ref{eq:Thm1})
    remains true, provided there is a generalized real-valued ground-state, that is, 
    a sequence $ \{ \hat\psi_{0,n} \}_{n \in \mathbb{N} } $ of real-valued 
    functions in the domain of $  H(\hat {b},\hat {v}) $    
    such that $ \langle \hat\psi_{0,n} , \, \hat\psi_{0,n} \rangle = 1 $ 
    for all $ n \in \mathbb{N} $ and  
    $ e_0(\hat {b}, \hat {v}) = 
    \inf_{n \in \mathbb{N}} \,  \langle \hat\psi_{0,n} , \, H(\hat {b},\hat {v}) \, \hat\psi_{0,n} \rangle $.
  \item 
    If $ \hat {b} $ is constant, one may use Lieb's inequality \cite[App.]{AHS78} 
    \begin{equation}
      e_0(\hat {b}, \hat {v}) \leq \case{1}{2} \, |\hat {b}| + e_0(0, \hat {v}) 
    \end{equation}
    to 
    further estimates the r.h.s.\ of (\ref{eq:Thm1}).
  \item
    The subsequent proof shows that analogous statements hold in more than two dimensions 
    and also for spinless (interacting) 
    many-particle systems obeying either Boltzmann or Bose statistics.
\end{itemize}
  If both $  \hat {b} $ and $  \hat {v} $ are centrally symmetric (or axially symmetric, when taking into account 
  the third dimension along the magnetic-field direction), a ground-state wave-function $ \hat\psi_0 $ of 
  $  H(\hat {b}, \hat {v} ) $, in the gauge (\ref{eq:Poncaregauge}), is real-valued 
  if and only if it is centrally symmetric or, equivalently, 
  an eigenfunction of $ L_3 $ with eigenvalue zero, $  L_3 \hat\psi_0 = 0 $.
  For vanishing magnetic field it is a standard textbook wisdom \cite{LL,GP1} that the ground state in
  a centrally symmetric potential is always given by a zero angular-momentum eigenfunction. 
  For non-vanishing magnetic fields this is wrong in general, cf.\ \cite{LC77,AHS78,SFI}. 
  However, in case the magnetic field $ \hat {b} > 0 $ is constant, the potential $\hat {v} $ is centrally symmetric and 
  the function $ | \bi{x} | \mapsto \hat {v}(|\bi{x}|) $ 
  is non-decreasing, then there exists, cf.\ \cite[Thm.~4.6]{AHS81}, at least 
  one $ \hat\psi_0 \neq 0 $ with $  L_3 \hat\psi_0 = 0 $ 
  in the (in general multi-dimensional) ground-state eigenspace 
  of 
  $ H(\hat {b},  \hat {v}) $ 
  and the additional assumption of Theorem~1 is fulfilled. 
  This applies for instance to the (centrally symmetric) harmonic-oscillator potential 
  \begin{equation}\label{def:osci}
    {v}_{\rm osc}(\bi{x}) := \frac{\omega^2}{2} \bi{x}^2, \qquad \omega \geq 0,
  \end{equation}
  and the attractive Coulomb-type potential,
  \begin{equation}
    {v}_{\rm cou}(\bi{x}) := - \frac{g}{\sqrt{\bi{x}^2 + \lambda^2}}, \qquad g,\, \lambda > 0.
  \end{equation}
  Accordingly, Theorem~\ref{Thm:1} implies the following upper bounds on the
  ground-state energy of a spinless charged particle in the plane $ \mathbb{R}^2 $ 
  subject to one of these potentials 
  and to a perpendicular,
  possibly inhomogeneous magnetic field $ {b} $ 
  the strength of which not exceeding a certain value, that is, 
  $ | {b}(\bi{x}) | \leq \hat {b} $ for all $ \bi{x} \in \mathbb{R}^2 $ with some
  constant $ \hat {b} > 0 $, 
  \begin{eqnarray}\label{eq:osci}
   & e_0({b},{v}_{\rm osc}) \leq e_0(\hat {b},{v}_{\rm osc}) = \sqrt{\big(\hat {b}/2\big)^2 + \omega^2},
  \\
  & e_0({b},{v}_{\rm cou}) \leq e_0(\hat {b},{v}_{\rm cou}).\label{eq:coul}
  \end{eqnarray}  
  The equality in (\ref{eq:osci}) follows from the explicitly known \cite{Foc28,Ma96} spectral 
  properties of $ H(\hat {b},{v}_{\rm osc}) $ with a constant $ \hat {b} $, 
  see also (\ref{eq:B+Osci}) below.

  Taking $ \omega = 0 $ and $ \hat {b} = \sup_{\bi{x} \in \mathbb{R}^2 } | {b}(\bi{x}) | $ in (\ref{eq:osci}) 
  yields the upper bound in the following sandwiching estimate
  on the ground-state energy of a spinless charged particle in a globally bounded and continuously differentiable 
  but otherwise arbitrary 
  magnetic field $ {b}: \mathbb{R}^2 \to \mathbb{R} $,   
  \begin{equation}\label{eq:MagnInCo}
    \case{1}{2} \, 
    \inf_{\bi{x} \in \mathbb{R}^2} \big| {b}(\bi{x}) \big|
    \leq e_0({b},0) \leq \case{1}{2} \, \sup_{\bi{x} \in \mathbb{R}^2} \big| {b}(\bi{x}) \big|.
  \end{equation}
  The lower bound in (\ref{eq:MagnInCo}) is a well-known consequence of the continuity of $ {b} $ together with  
  the non-negativity of $ H({b}, 0 ) $ and  
  that of its associated Pauli operator \cite{CFKS}, in symbols, $ H({b},0) \geq 0 $ and $ H({b}, \pm {b}/2 )\geq 0 $. \\

We close this section with a simple

\begin{proof}[\bf Proof of Theorem~\ref{Thm:1}]
The vector potentials $ \bi{{a}} $ and $ \hat\bi{ {a}} $ defined by (\ref{eq:Poncaregauge}) satisfy 
the pointwise inequality 
\begin{equation}\label{eq:boundA}
\fl
   |\bi{{a}}(\bi{x}) |  \leq | \bi{x} | \int_0^1 d \xi \, \xi \; | {b}(\xi \bi{x}) | \leq
 | \bi{x} | \int_0^1 d\xi \, \, \xi \;  \hat {b} (\xi \bi{x})  
 = |  \hat\bi{ {a}}(\bi{x}) |
\end{equation}
by the triangle inequality and the assumption $  | {b}(\bi{x}) |\leq \hat {b}(\bi{x})  $ for all $ \bi{x} \in \mathbb{R}^2 $.
The non-commutative binomial formula gives
\begin{eqnarray}\label{eq:binomial}
\fl
  \langle \psi , \, H(\hat {b}, \hat {v}) \, \psi \rangle  = &
   \langle \psi , \, H({b}, {v}) \, \psi \rangle + {\rm Re } \, \langle \psi , ( \bi{{a}}(\bi{Q}) - \hat\bi{{a}}(\bi{Q}) ) \cdot \bi{P}  \, \psi \rangle \nonumber \\
  &  
  +   \case{1}{2} \, \langle \psi , ( \hat\bi{{a}}(\bi{Q})^2 - {\bi{{a}}}(\bi{Q})^2 ) \, \psi \rangle
  +   \langle \psi , (\hat{v}(\bi{Q}) - {v}(\bi{Q}) ) \, \psi \rangle 
\end{eqnarray}
for any $ \psi $ in the domain of $ H(0,0) $, hence in those of $  H({b},{v}) $ and $  H(\hat {b},\hat {v}) $
by the Kato-Rellich theorem \cite{CFKS}. 
The real part of $ \langle \psi , ( \bi{{a}}(\bi{Q}) - \hat\bi{{a}}(\bi{Q}) ) \cdot \bi{P}  \, \psi \rangle $ 
on the r.h.s.\ vanishes if $ \psi $ is real-valued, because $ \bi{P} $ acts as the gradient divided by the imaginary 
unit $ i = \sqrt{-1} $. 
Employing (\ref{eq:boundA}) and the assumption 
on the potentials the last two terms on the r.h.s.\ of (\ref{eq:binomial}) are both seen to be non-negative.
Altogether this yields 
$
  e_0({b},{v}) \leq \langle \hat\psi_0 \, ,  H({b}, {v}) \, \hat\psi_0 \rangle \leq  
\langle \hat\psi_0 \, , H(\hat {b}, \hat {v}) \, \hat\psi_0 \rangle 
=  e_0(\hat {b}, \hat {v}) 
$
where the first inequality is a
consequence of the Rayleigh-Ritz variational principle.
\end{proof}

\section{Monotonicity of density matrices}
Diamagnetic monotonicity of density matrices (or only of their diagonals) 
implies  monotonicity of the corresponding ground-state energies in the zero-temperature limit.
Therefore, Facts~1 and 2 show that the former monotonicity can also neither hold in the presence of general potentials for
two homogeneous magnetic fields nor in case of vanishing potential for two general magnetic fields. 
Accordingly, one can hope to find a diamagnetic monotonicity of density matrices only under additional assumptions 
on the magnetic fields and/or the potentials.

\subsection{Globally constant magnetic fields and harmonic-oscillator potential}
In case of a homogeneous, that is, globally constant magnetic field $ {b} \in \mathbb{R} $ and 
the harmonic-oscillator potential (\ref{def:osci}), 
the density matrix is exactly known and, in the gauge (\ref{eq:Poncaregauge}),
explicitly given \cite{Pa71,Ma96} in terms of the ground-state energy 
$ \Omega_{b}:= \sqrt{ ({b}/2)^2 + \omega^2} $, cf. (\ref{eq:osci}), and hyperbolic functions by the Mehler-type of formula \cite{CFKS}
\begin{eqnarray}\label{eq:B+Osci}
\fl
 \langle \bi{x} |  \e^{- \beta H({b}, {v}_{\rm osc})}  | \bi{y} \rangle = & \frac{\Omega_{b}}{2 \pi \sinh\left(\beta \Omega_{b} \right)} 
 \exp\!\left\{ - \frac{\Omega_{b}}{2  \tanh\left(\beta \Omega_{b} \right)} \! \left[ \bi{x}^2 + \bi{y}^2 \! -   2  \bi{x} \cdot \bi{y}  \frac{\cosh(\beta {b} /2)}{  \cosh\left(\beta \Omega_{b} \right) } \right]   \right\} \nonumber \\
 &  \times \exp\Bigg\{   
 i \, \Omega_{b} \, \frac{\sinh(\beta {b} / 2)}{\sinh(\beta \Omega_{b} )}  \big( x_2 y_1 -  x_1 y_2  \big)    \Bigg\}.
\end{eqnarray}

By discriminating the cases $ \bi{x} \cdot \bi{y} \geq 0 $ and $ \bi{x} \cdot \bi{y} < 0 $, an 
elementary (but somewhat tedious) 
calculation shows that the function $ | {b} | \mapsto 
 | \langle \bi{x} | \exp[ -\beta H( {b} , {v}_{\rm osc} ) ] | \bi{y} \rangle | $ is
non-increasing for all $ \omega \geq 0 $, all 
$  \bi{x} $, $ \bi{y} \in \mathbb{R}^2 $ and all $ \beta > 0 $. 
This monotonicity extends (\ref{eq:osci}) in case of a globally constant $ {b} $ and, in view of Fact~1, 
is a particularity of the harmonic oscillator. 

As for the monotonicity in the potential, we remark the following. 
The function 
$ \omega \mapsto \langle \bi{x} | \, \exp[ - \beta H( {b} , {v}_{\rm osc} )] \, | \bi{x} \rangle $ is obviously 
non-increasing for all $ b \in \mathbb{R} $, all $  \bi{x} \in \mathbb{R}^2 $ and all $ \beta > 0 $. 
This property of the harmonic-oscillator density-matrix diagonal is a stronger one than 
the universally valid (reverse) 
monotonicity of the free energy 
$ - \beta^{-1} \ln  \int_{\mathbb{R}^2} d^2 x \,  
\langle \bi{x} | \, \exp[ - \beta H( {b} , {v} )] \, | \bi{x} \rangle  $ 
when $ {v} $ is pointwise increased, cf.\ the Introduction.
However, the harmonic oscillator already illustrates that, 
in contrast to the situation with $ {b} = 0 $, monotonicity of 
the density-matrix off-diagonal in the potential in general ceases 
to hold for $  {b} \neq 0 $. More precisely, by elementary calculations one finds
\begin{description}
  \item{\bf Fact 3:}  There exist two constants $ {b} $, $ \beta > 0 $ and a non-zero 
    $ \bi{x} \in \mathbb{R}^2 $, such that the function 
    $ \omega \mapsto | \langle \bi{x} | \, \exp[ - \beta H( {b} , {v}_{\rm osc} )] \, | - \bi{x} \rangle | $
    is increasing in some neighbourhood of some $ \omega_0 > 0 $.
\end{description}

\subsection{Non-constant magnetic fields and vanishing potential}
Loss and Thaller \cite{LT97} studied the density matrix  
associated with an inhomogeneous magnetic field $ \hat {b} $ which is globally bounded from below by a non-negative 
constant one. 
In particular, they have shown \cite[Thm.~1.3]{LT97} that the inequality 
\begin{equation}\label{eq:LT}
  \Big| \langle \bi{x} | \, \e^{- \beta H(\hat {b} , 0 ) } \, | \bi{y} \rangle \Big| \leq \frac{{b}}{4 \pi \sinh(\beta {b}/2)} \, 
   \exp\Big[ - \frac{(\bi{x} - \bi{y})^2}{2 \beta} \Big]
\end{equation}
holds for all $ \bi{x} $, $ \bi{y} \in \mathbb{R}^2 $ and all $ \beta > 0 $ 
as long as $ {b} \leq \hat {b}(\bi{x}) $ for all $ \bi{x} \in \mathbb{R}^2 $ with some constant $ {b} \geq 0 $. 
The Gaussian in (\ref{eq:LT}) coincides
with that of the free density matrix $ \langle \bi{x} | \exp[ - \beta H( 0 , 0 ) ] \, | \bi{y} \rangle $
and the pre-factor with the diagonal $ \langle \bi{x} | \exp[ - \beta H( {b} , 0 )] \, | \bi{x} \rangle $,
which is actually independent of $ \bi{x} \in \mathbb{R}^2 $, see (\ref{eq:B+Osci}) with $ \omega = 0 $. 

Erd\H{o}s \cite{E97} has shown that (\ref{eq:LT}) provides basically the best upper bound on the 
density-matrix off-diagonal one can hope for, unless $ \hat {b} : \mathbb{R}^2 \to \mathbb{R} $
has further properties. 
More precisely, he proved that (\ref{eq:LT}) cannot be improved universally by replacing its r.h.s.\
by $ | \langle \bi{x} | \exp[ - \beta H({b},0) ] | \bi{y} \rangle | $, which has more rapid Gaussian decay 
as $ | \bi{x} - \bi{y} | \to \infty $, cf. (\ref{eq:B+Osci}).
\begin{description}
  \item{{\bf Fact 4} (\cite[II.16]{E97}):} 
    Define $ \hat {b}(\bi{x}) :=  (1 + x_1^2/ \lambda^2 ) \, {b} $  with two constants $ {b} $, $ \lambda > 0 $. Then
    $ {b} \leq \hat {b}(\bi{x}) $  for all $ \bi{x} \in \mathbb{R}^2 $, but  
    $ | \langle \bi{x} | \exp[ - \beta H(\hat {b} , 0 ) ] | \bi{y} \rangle | 
    >  | \langle \bi{x} | \exp[ - \beta H({b} , 0 ) ] | \bi{y} \rangle | $ for  
    some $ \lambda > 0 $, some $  \bi{x} $, $ \bi{y} \in \mathbb{R}^2 $ and some $ \beta > 0 $. 
\end{description}
Nevertheless, 
in the next subsection it will turn out that (\ref{eq:LT}) 
can be improved at the cost of allowing only a restricted class 
of inhomogeneous magnetic fields (including the one of Fact~4).

\subsection{Magnetic fields and potentials which are constant along one direction}\label{Sec:Iwa}
In this subsection we will restrict ourselves to the special class of magnetic fields $ {b} $ 
which do not depend on the second
co-ordinate $ x_2 $. This class covers the case of globally constant fields, for which the spectrum of the 
Schr\"odinger operator $ H({b}, 0) $ is well known \cite{Foc28,Lan30,GP2,Ma96} to consist only of 
isolated harmonic-oscillator like eigenvalues $ (n - 1/2) | {b} | $, $ n \in \mathbb{N} $, of infinite degeneracy, 
the so-called Landau levels. 
However, in case such a field is not globally constant, 
$ H({b}, 0) $ is conjectured \cite{CFKS}
to possess only (absolutely) continuous spectrum. 
The first rigorous proof of this conjecture was given by 
Iwatsuka \cite{I85} under certain additional assumptions, 
see also \cite{CFKS,MP97}. In acknowledgement of this achievement, 
for such fields $ H( {b}, 0) $  
often goes under the name \emph{Iwatsuka model}. 

In what follows, it is most convenient to choose the asymmetric gauge defined by
\begin{equation}\label{eq:asymgauge}
  {a}_1(\bi{x}) := 0, \qquad {a}_2(x_1) := \int_0^{x_1} \! d x_1' \, {b}(x_1'). 
\end{equation}
Then the resulting Schr\"odinger operator 
\begin{equation}\label{eq:IwatsukaH}
  H( {b}, {v}) = \case{1}{2} P_1^2 +  \case{1}{2} \big(P_2 - {a}_2(Q_1)\big)^2 + {v}(Q_1),
\end{equation} 
is translation invariant along the $ x_2 $-direction, provided the potential $ {v} $ 
does not depend on $ x_2 $, too. The operator (\ref{eq:IwatsukaH}) can therefore 
be decomposed by partial Fourier transformation into the one-parameter family 
$  H_1(k) :=  P_1^2/2 + (k - {a}_{2}(Q_1) )^2/2 + {v}(Q_1) $, $ k \in \mathbb{R} $,
of Schr\"odinger operators for the $ x_1 $-direction. 
As a consequence, one obtains for the density matrix of (\ref{eq:IwatsukaH})
\begin{eqnarray}
\fl
\langle \bi{x} | \, \e^{- \beta H({b}, {v})} \, | \bi{y} \rangle  & = \int_\mathbb{R}  \frac{d k}{2 \pi} \; 
                    \langle x_1 | \, \e^{-\beta H_1(k) } \, | y_1 \rangle    \; \e^{i  (x_2 - y_2 )  k } \label{eq:eineR}\\
                    & = (2\pi \beta)^{-1} 
                     \exp\!\Big[- \frac{(\bi{x}- \bi{y})^2}{2\beta} \Big] \label{eq:IwaPa} \\
        & \mkern-35mu
        \times \int\mathbb{P}_{x_1,y_1}^{\, 0 \, , \, \beta}\big(d w\big) \, \exp\!\Big[ \, i\, (x_2 - y_2 ) \, \mu_\beta({a}_{2}\circ w ) 
       -\frac{\beta}{2}  \sigma_\beta^2({a}_{2}\circ w )  -  \beta \mu_\beta( {v} \circ w) \Big]. \nonumber
\end{eqnarray}
For the derivation of the second equality we used the Feynman-Kac formula \cite{SFI,R}
for the density matrix of $ H_1(k) $ in (\ref{eq:eineR}), which involves path integration with 
respect to the probability measure $ \mathbb{P}_{x_1,y_1}^{\, 0\, , \, \beta} $ of the 
one-dimensional Brownian bridge
going from $ x_1 $ at time $ 0 $ to $ y_1 $ at time $ \beta $. 
The latter is 
the unique normalized Gaussian measure on the set of continuous paths $ w: [0, \beta ] \to \mathbb{R} $, 
$ \tau \mapsto w(\tau) $
with mean function $ \tau \mapsto x_1 + (y_1 -x_1 ) \tau /\beta $ and covariance function 
$ (\tau , \tau' ) \mapsto \min\{  \tau , \tau' \} - \tau \tau' / \beta $.
%
In (\ref{eq:IwaPa}) we are making use of the notations
\begin{eqnarray}
\mu_\beta(f \circ w) & := \beta^{-1} \int_0^\beta d \tau \, f\big(w(\tau)\big),\\
\sigma_\beta^2(f \circ w) & := \mu_\beta\big( (f \circ w)^2 \big) - \big( \mu_\beta(f \circ w) \big)^2 \geq 0 \label{def:sigma}
\end{eqnarray}
for the mean and variance of the composition $ f \circ w $ of 
a continuous function $ f : \mathbb{R} \to \mathbb{R} $ 
and a path $ w $ with respect to the uniform time average $ \beta^{-1} \int_0^\beta d \tau \, ( \cdot )  $. 
To obtain (\ref{eq:IwaPa}) we also interchanged the (Lebesgue) integration with respect to $ k $ 
and the Browian-bridge integration by referring to the Fubini-Tonelli theorem.  
Thanks to translation invariance along the $ x_2 $-direction the density matrix of (\ref{eq:IwatsukaH}) 
depends on $ x_2 $ and $ y_2 $ only through their difference $ x_2 - y_2 $. 
Moreover, a nice feature of the path-integral representation (\ref{eq:IwaPa}) is that its 
integrand contains the magnetic field but
is nevertheless non-negative if $ x_2 = y_2 $. 
This enables one to estimate the density matrix of (\ref{eq:IwatsukaH})  
by using standard inequalities of general integration theory (cf.\ \cite{LiLo})
without loosing the dependence on the magnetic field.
For example, the triangle inequality applied to (\ref{eq:IwaPa}) gives the (gauge-independent) estimate  
\begin{equation}\label{eq:IwaE1}
\fl
  \Big| \big\langle \bi{x} | \, \e^{- \beta H({b}, {v})} \, | \bi{y}  \big\rangle  \Big| 
  \leq \Big| \big\langle (x_1,0) | \, \e^{- \beta H({b}, {v})} \, | (y_1,0)  \big\rangle   \Big| 
  \, \exp\!\Big[-\frac{(x_2 - y_2)^2}{2 \beta}\Big],
\end{equation}
where, in the chosen gauge (\ref{eq:asymgauge}), it would not be necessary to take the absolute value of the r.h.s.. 
The following theorem provides a generalization of (\ref{eq:IwaE1}). For an interesting application of (\ref{eq:IwaPa})
in case $ a_2 $ is a Gaussian random field and $ v = 0 $, see \cite{U00}.
\begin{theorem}\label{Thm2}
 Let $  {b} $, $  \hat {b} $ and $ {v} $, $ \hat {v} $ be two magnetic fields and two potentials, 
 all four of them not depending on the second co-ordinate $ x_2 $. 
 Then the pointwise inequalities $ | {b}(x_1) | \leq  \hat {b} (x_1) $ 
 and $  {v}(x_1) \leq \hat {v}(x_1) $ for all $ x_1 \in \mathbb{R} $ imply the inequality
 \begin{equation}\label{eq:Thm2}
 \fl
   \Big| \big\langle \bi{x} | \, \e^{- \beta H( \hat {b},  \hat {v})} \, | \bi{y}  \big\rangle \Big|
   \leq  \Big| \big\langle (x_1, 0) | \,  \e^{- \beta H( {b}, {v})} \, |  (y_1, 0)  \big\rangle \Big| 
   \, \exp\!\Big[-\frac{(x_2 - y_2)^2}{2 \beta}\Big]  
 \end{equation}
 for all $ \bi{x} = (x_1, x_2) \in  \mathbb{R}^2 $, all $ \bi{y} = (y_1, y_2) \in  \mathbb{R}^2 $ and all $ \beta > 0 $.
\end{theorem}

\noindent
Three special cases of (\ref{eq:Thm2}) are worth to be mentioned separately:
\begin{itemize}
  \item
    Theorem~\ref{Thm2} and translation invariance along the $ x_2 $-direction imply monotonicity
    of the density-matrix diagonal in the magnetic field and the potential in the sense that
    $
      \big\langle \bi{x} | \, \exp[- \beta H( \hat {b},  \hat {v})] \, | \bi{x}  \big\rangle 
     \leq \big\langle \bi{x} | \, \exp[- \beta H( {b},  {v})] \, | \bi{x}  \big\rangle  
    $ 
    for all $ \bi{x}\in  \mathbb{R}^2 $. In particular, in the zero-temperature limit $ \beta \to \infty $ the monotonicity 
    $ e_0({b},{v}) \leq  e_0(\hat {b}, \hat {v}) $
    of the corresponding ground-state energies emerges.
    In general, this monotonicity  is not covered by Theorem~\ref{Thm:1} as the case $ {b} = 0 $ already illustrates.
  \item
    If $ {b} \neq  0 $ is globally constant and $ {v} = \hat {v} = 0 $, (\ref{eq:Thm2}) together with (\ref{eq:B+Osci}) 
    yields  
    the following improvement of (\ref{eq:LT}) for the present situation, 
    in which $ \hat {b} $ does not depend on $ x_2 $,
    \begin{equation}\label{eq:LTErsatz}
      \fl
      \Big| \big\langle \bi{x} | \, \e^{- \beta H( \hat {b},  0)} \, | \bi{y}  \big\rangle \Big|
      \leq \frac{{b}}{4 \pi \sinh( \beta {b} /2) } \, \exp\Big[ - \frac{{b}}{4}
      \frac{ (x_1 - y_1)^2}{\tanh(\beta {b} /2)}  - \frac{(x_2 - y_2)^2}{2 \beta} \Big].
    \end{equation}
    The Gaussian decay on the r.h.s.\ of (\ref{eq:LT}) as $ | x_1 - y_1 | \to \infty $ 
    may thus be replaced by that 
    of $ | \big\langle \bi{x} | \exp\left[ - \beta H({b},0) \right] | \bi{y}  \big\rangle | $. However,  
    as illustrated by Fact~4, this is not allowed for the Gaussian decay along the perpendicular direction in the plane. 
    In this sense (\ref{eq:LTErsatz}) is optimal for the present situation. For given non-negative 
    $ \hat {b}: \mathbb{R} \to [0, \infty[ $,
    the r.h.s. of (\ref{eq:LTErsatz}) is minimized by taking $ {b} = \inf_{x_1 \in \mathbb{R}} \hat {b}(x_1) $.
    \item
      If $ \hat {b} > 0 $ is globally constant and $ {v} = \hat {v} = 0 $, (\ref{eq:Thm2}) together with (\ref{eq:B+Osci}) 
      yields the following lower estimate 
      on certain density-matrix off-diagonals: 
      \begin{equation}\label{eq:LTErsatz2}
        \fl
        \frac{\hat {b}}{4 \pi \sinh\big(\beta \hat {b}/2\big)} \, 
        \exp\Bigg[- \frac{\hat {b}}{\stackrel{\mbox{}}{4}} \frac{(x_1 - y_1)^2}{\tanh\big(\beta \hat {b}/2 \big)} \Bigg] 
        \leq \Big| \big\langle (x_1, 0) \, \big| \, \e^{- \beta H( {b},  0)} \, \big| \, (y_1, 0)  \big\rangle \Big|. 
      \end{equation}
      For given $ {b}: \mathbb{R} \to \mathbb{R} $,
      the l.h.s. of (\ref{eq:LTErsatz2}) is maximized by taking $ \hat {b} = \sup_{x_1 \in \mathbb{R}} | {b}(x_1) | $.
\end{itemize}
\begin{proof}[\bf Proof of Theorem~\ref{Thm2}] 
  Without loosing generality, we may choose the gauge (\ref{eq:asymgauge}) for $ {b} $ and analogously for $  \hat {b} $.
  Then we observe the pathwise non-negativity
  \begin{eqnarray}
  \fl 
  \sigma_\beta^2(\hat {a}_{2}\circ w)  - \sigma_\beta^2( {a}_{2}\circ w ) \nonumber \\
  \mkern-100mu\lo{=}  \frac{1}{2 \beta^{2}} \int_0^\beta \! d \tau  \int_0^\beta \! d \tau' \, \Big\{ 
  \big[ \hat {a}_{2}(w(\tau)) - \hat {a}_{2}(w(\tau')) \big]^2 
       - \big[ {a}_{2}(w(\tau)) - {a}_{2}(w(\tau')) \big]^2  \Big\}
       \nonumber \\
  \mkern-100mu \lo{=}  \frac{1}{2 \beta^{2}} \int_0^\beta \! d \tau  \int_0^\beta \! d \tau' \, 
       \Big[ {a}_+\big(w(\tau)\big) - {a}_+\big(w(\tau')\big) \Big] \,
     \Big[ {a}_-\big(w(\tau)\big) - {a}_-\big(w(\tau')\big)\Big] \geq 0. \label{eq:binom}
  \end{eqnarray}
  Here the non-negativity of the last integrand follows from the fact that the two functions
  $ x_1 \mapsto {a}_\pm(x_1) := \hat {a}_{2}(x_1) \pm  {a}_{2}(x_1) = \int_0^{x_1} d x_1' \, 
  \big[ \hat {b}( x_1') \pm {b}( x_1') \big] $, cf. (\ref{eq:asymgauge}), 
  are both non-decreasing
  since $ \hat {b}(x_1') \geq | {b}(x_1') | \geq \mp {b}(x_1') $ for all $ x_1' \in \mathbb{R} $ by assumption. 
  Besides (\ref{eq:binom}) we have $ \mu_\beta(v \circ w) \leq \mu_\beta(\hat v \circ w) $, also by assumption.
  Employing both inequalities in (\ref{eq:IwaPa}) with $ x_2 = y_2 $, 
  the proof is completed with the help of (\ref{eq:IwaE1}) (with $ {b} $ replaced by $ \hat {b} $).
\end{proof}

\section{Concluding remark}
We conclude by specifying part of the open problem \cite[II.12]{E97} which deals with the complementary situation of (\ref{eq:LT}).
\begin{description}
  \item{\bf Open problem:} Let $ {b}: \mathbb{R}^2 \to \mathbb{R} $ be continuously differentiable and
    satisfy $ | {b}(\bi{x}) | \leq \hat {b} $ for all $ \bi{x} \in \mathbb{R}^2 $ with some constant $ \hat {b} > 0 $.
    Prove or disprove the assertion
    \begin{equation}\label{eq:openp}
         \Big| \big\langle \bi{x} \, \big| \, \e^{- \beta H( \hat {b},  0)} \, \big| \, \bi{y}  \big\rangle \Big|
         \leq \Big| \big\langle \bi{x} \, \big| \, \e^{- \beta H( {b},  0)} \, \big| \, \bi{y}  \big\rangle \Big| 
      \end{equation}
      for all $ \bi{x} $, $  \bi{y} \in \mathbb{R}^2 $ and all $ \beta > 0 $.
\end{description}
The present note contains partial support for the validity of this assertion. 
Namely, (\ref{eq:openp}) is true at least in each of the following three limiting cases:
\begin{itemize}
  \item
    to logarithmic accuracy in the zero-temperature limit, by the second inequality in (\ref{eq:MagnInCo}).
  \item
    if $ {b} $ is constant along the $ x_2 $-direction and $ x_2 = y_2 $, by (\ref{eq:LTErsatz2}).
  \item
    if $ {b} $ is globally constant, by (\ref{eq:B+Osci}) with $ \omega = 0 $.
\end{itemize}
The third case is well known.

\ack
This work was partially supported by the Deutsche Forschungsgemeinschaft
under grant no.\ Le 330/12. The latter is a project within the
Schwerpunktprogramm ``Inter\-agierende stochastische Systeme 
von hoher Komplexit\"at'' (DFG Priority Programme SPP 1033).

\maketitle

\end{document}